\documentclass[aps,prl,twocolumn,secnumarabic,nobalancelastpage,amsmath,amssymb,floatfix,showpacs,superscriptaddress,groupedaddress,longbibliography,reprint]{revtex4-1}

\usepackage{graphicx}
\usepackage[T1]{fontenc}
\usepackage[USenglish]{babel}
\addto\captionsUSenglish{}

\usepackage{hyperref}

\DeclareMathOperator{\Tr}{Tr}

\begin{document}
\title{Single-photon interference due to motion in an atomic collective excitation}
\author{D.~J.~Whiting}\email{daniel.whiting@durham.ac.uk}
\author{N.~\v{S}ibali\'{c}}
\author{J.~Keaveney}
\author{C.~S.~Adams}
\author{I.~G.~Hughes}
\affiliation{Joint Quantum Center (JQC) Durham-Newcastle, Durham University, Department of Physics, South Road, Durham, DH1 3LE, United Kingdom}

\date{\today}

\begin{abstract}
We experimentally demonstrate the generation of heralded bi-chromatic single photons from an atomic collective spin excitation (CSE). The photon arrival times display collective quantum beats, a novel interference effect resulting from the relative motion of atoms in the CSE. A combination of velocity-selective excitation with strong laser dressing and the addition of a magnetic field allows for exquisite control of this collective beat phenomenon. The present experiment uses a diamond scheme with near-IR photons that can be extended to include telecommunications-wavelengths or modified to allow storage and retrieval in an inverted-Y scheme.
\end{abstract}

\maketitle

\textit{Introduction} --- 
Quantum-state engineering is of critical importance to the development of quantum technologies. Atomic media are an attractive option for realising these technologies~\cite{Hammerer2010}, providing well-defined optical transitions, long coherence times, frequency-matched high-brightness single-photon sources \cite{MacRae2012a,Chou2004}, quantum memories  \cite{Saunders2016,Hosseini2011,Lvovsky2009} and repeaters \cite{Duan2001}, coherent control protocols based on slow-light and adiabatic following \cite{Fleischhauer2000}, and strong non-linearities that produce controllable phase shifts \cite{Pritchard2013}.
While there are clear advantages over solid-state approaches~\cite{Ladd2010}, the technological complexity of typical cold-atom experiments presents a challenge for scaling and wider application.

In contrast to cold-atom systems, thermal atomic vapor experiments provide a reproducible and scalable hardware platform.
Their use has enabled the development of many practical devices including chip-scale atomic clocks~\cite{Kitching2002}, brain sensors~\cite{Sander2012} and microwave electrometers~\cite{Sedlacek2012}.
However, the inability to address individual atomic states in a controlled manner, due to multi-level degeneracy and motional broadening, inhibits their wider use in quantum state engineering applications.
Optical pumping is conventionally used for initial state preparation and buffer gases~\cite{Brandt1997} and anti-relaxation coatings~\cite{Budker2007,Balabas2010} can be employed to mitigate decoherence processes for ground state atoms.
However, for schemes involving excited states~\cite{Sargsyan2010} or thin-cells~\cite{Sargsyan2016} these methods usually cannot be applied.
An alternative solution is to apply a strong magnetic field that resolves the multi-level degeneracy.
This method has recently been shown to simplify non-linear atom-light interactions in thermal vapors, resulting in enhanced control of electromagnetically induced transparency~\cite{Whiting2016} and absorption~\cite{Whiting2015}.
Another major challenge facing the application of thermal atomic vapors to quantum state engineering is motion-induced dephasing~\cite{Firstenberg2013}, because of the broad atomic velocity distribution.
It is therefore interesting to consider novel quantum states that exploit this motion, for example when a single excitation is stored in an entangled state of two atoms with relative motion.
This state was discussed theoretically in the 1970's, but was deemed "impossible to observe directly" \cite{Haroche1976} in thermal vapors due to the wide spread of velocities rapidly washing out the spatial correlations between atoms.

In this letter we demonstrate a method to engineer this type of collective state in a thermal atomic vapor.
The prepared state consists of a single excitation as a robust collective superposition of two velocity classes, whose coherent nature is demonstrated by measuring collective quantum beats~\cite{Haroche1976}.
The single excitation is emitted as a \emph{single photon with two frequencies}. At present there is much interest in these `bi-chromatic' photons as they could be used to entangle spatially separated quantum memories or perform spectroscopy with small numbers of photons ~\cite{Clemmen2016,Treutlein2016}.
Combining the application of a large magnetic field and strong laser-dressing in a velocity selective ladder-type excitation, we demonstrate excellent control over the state preparation.

\textit{State preparation} --- 
During the state preparation a strong magnetic field allows individual control over the internal atomic states, and a ladder-type excitation with strong laser dressing allows tunable selection of the external (motional) states.
A magnetic field ($B=0.6$~T) splits the atomic states according to their projection of spin-orbit coupling $m_{\rm J}$, by energy $m_{\rm J} \mu_{\rm B} B$, where $\mu_{\rm B}$ is the Bohr magneton.
This field, provided by permanent neodymium magnets, separates the optical transitions of the atom by more than their Doppler-broadened linewidth~\cite{Weller2012,Zentile2014}.
A pump laser can then be tuned to address only those atoms from the ensemble that are in the chosen $m_{\rm J}$ state, $|\rm g\rangle$, reducing the internal degrees of freedom of the system to four coupled levels [Fig.~\ref{fig1}(a) inset].
\begin{figure*}
	\includegraphics[width=\textwidth]{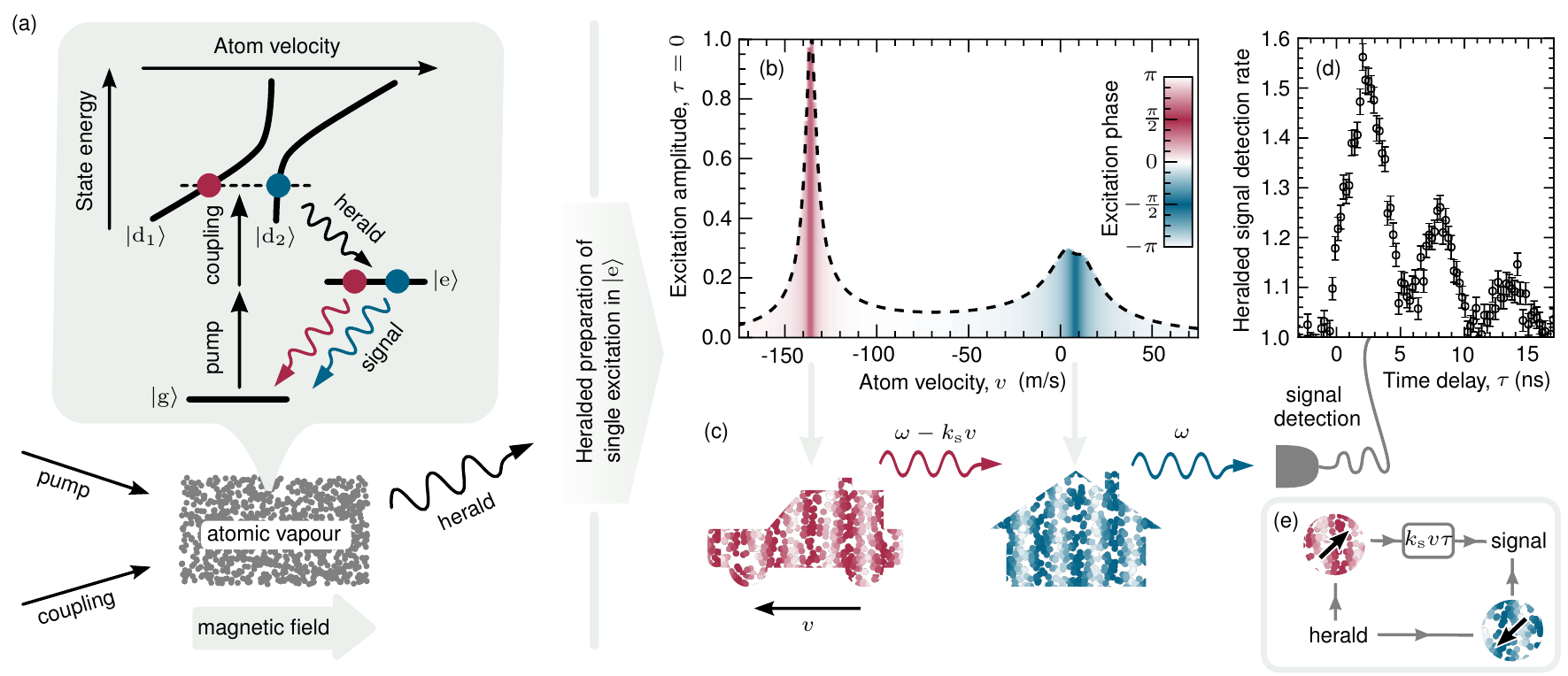}
	\caption{\label{fig1}Preparation of a single excitation as a collective superposition of two atomic velocity groups.
		(a) A thermal vapor of $^{87}$Rb is continuously driven by weak pump- and strong coupling-laser beams.
		A strong magnetic field of 0.6~T, applied with permanent magnets, simplifies the internal level structure by isolating only the four levels shown in the inset (in the semi-dressed picture).
		The coupling laser dresses the atoms, so that atoms with two different velocities (shown in red and blue in inset) are preferentially excited.
		Detection of a "herald" photon heralds the preparation of a single excitation in the level $|\rm e\rangle$ in the form of a spin-wave.
		(b) The excitation is mostly split amongst two velocity classes, one stationary and one moving away from the signal detector in (c), with an initial phase difference of $\pi$.
		Due to the Doppler effect, light emitted from these two classes of atoms will be shifted in frequency causing beats in the signal photon detection (d).
		The beats demonstrate that this set-up forms an interferometer (e), where detection of a herald photon coherently splits a single excitation and stores it in atoms moving at two different velocities, before recovering the excitation in a common signal channel.}
\end{figure*}
A ladder-type excitation scheme with co-propagating pump and coupling lasers [Fig.~\ref{fig1}(a)] selects a narrow group of resonant atoms from the broad velocity distribution.
A strong coupling-laser dresses the bare atomic states, $|\rm a\rangle$ and $|\rm b\rangle$, allowing simultaneous excitation of two narrow velocity-groups (with well defined phases) satisfying the condition $2k v_z = \frac{1}{2}(\Delta_{\rm c} \pm \sqrt{\Delta_{\rm c}^2+2\Omega_{\rm c}^2})$ (Fig.~\ref{fig1}).
These two groups correspond to the dressed states $|\rm d_1\rangle$ and $|\rm d_2\rangle$ in Fig.~\ref{fig1}(a) inset.
By choosing the detuning $\Delta_{\rm c}$ and driving strength $\Omega_{\rm c}$ of the coupling laser, with wavevector $k$, one can set the velocities $v_z$ of the two excited velocity classes.
For a negatively (red) detuned coupling laser, these correspond to one nearly stationary group and one moving away from the detector [Fig.~\ref{fig1}(b)].

A single collective excitation is produced by heralding on the spontaneous decay of the excited atoms.
The herald photon maps the instantaneous relative phase of the atoms [Fig.~\ref{fig1}(b)], from the steady state under strong laser driving, into the excited state $|\rm e\rangle$.
Since the strong driving preferentially selects two atomic velocity classes, the photon detection heralds coherent splitting of the single excitation into these two velocity groups.
The driving lasers and the herald and signal output channels fulfil the wave-matching condition as in usual diamond four-wave mixing schemes \cite{Willis2010a,Srivathsan2013}.
Due to this, the single excitation takes the form of a spin-wave, picking out a preferential output direction for collective emission of the signal photon~\cite{Zhao2009}.
Because of the atomic motion, the emission from the moving group of atoms will be Doppler shifted with respect to that of the stationary atoms [car and house in Fig.~\ref{fig1}(c)].
This frequency shift leads to interference and the observation of beats in the signal emission [Fig.~\ref{fig1}(d)], demonstrating the persistence of coherence in the single excitation split between two velocity groups.
In contrast to usual quantum beats, that originate due to state superposition within the single atom structure~\cite{Haroche1973,Aspect1984,Wade2014}, these beats originate due to a superposition of atoms with different velocities being in the same internal excited state $|\rm e\rangle$ [Fig.~\ref{fig1}(e)].
Beating of light fields emitted by two groups of atoms with different velocities has previously been observed in superradiant emission from thermal ensembles after pulsed excitation~\cite{Gross1978}.
However these superradiant beats cannot be observed on a single photon level, since which-path information is stored in the excited state regarding which atoms decay in the process; one could in principle check, for each emitted photon, which velocity class is in the excited state.
Finally, we note that single photon beats can be observed in cold atoms (only one velocity class) by using an additional laser to dress the levels involved~\cite{Du2008,Yang2015}.

\textit{Experimental details} --- 
Experimentally we use $^{87}$Rb atoms in a diamond scheme with energy levels denoted $|g\rangle = 5\mathrm{S}_{1/2}(m_{J}=1/2)$, $|a\rangle =  5\mathrm{P}_{3/2}(m_{J}=3/2)$, $|b\rangle = 5\mathrm{D}_{3/2}(m_{J}=1/2)$ and $|e\rangle = 5\mathrm{P}_{1/2}(m_{J}=-1/2)$.
Continuous-wave pump and coupling fields are tuned to the $|g\rangle\rightarrow|a\rangle$ and $|a\rangle\rightarrow|b\rangle$ resonances at 780~nm and 776~nm respectively.
The pump and coupling fields, with angular separation 10~mrad, are focused to $50~\mu$m (1/e$^{2}$ waists) and overlapped at the center of a 2~mm long atomic vapor cell.
The cell, containing rubidium (isotopic abundance 98\% $^{87}$Rb and 2\% $^{85}$Rb), is heated to 90$^{\circ}$C.
The cell also contains buffer gasses which contribute an additional broadening of 7~MHz to the 5S$\rightarrow$5P transitions and 13~MHz to the 5P$\rightarrow$5D transitions.
The pump and coupling powers are $4~\mu$W and 40~mW respectively, which correspond approximately to Rabi frequencies of $\Omega_{\rm p}/2\pi=35$~MHz and $\Omega_{\rm c}/2\pi=280$~MHz.
The herald and signal photons are spontaneously emitted on the transitions $5\mathrm{D}_{3/2}(m_{J}=1/2)\rightarrow 5\mathrm{P}_{1/2}(m_{J}=-1/2)\rightarrow 5\mathrm{S}_{1/2}(m_{J}=1/2)$ at 762~nm and 795~nm respectively.
In this configuration the generated photons are emitted in the forward direction to fulfil the phase matching criterion $\bf k_{\rm p}+ k_{\rm c}= k_{\rm h}+ k_{\rm s}$.
After being separated from the pump light by narrowband interference filters and polarisation filtering (see supplemental material) the generated photons are collected into single mode optical fibres and detected by avalanche photodiodes.
A timing card with a 27~ps resolution records the photon detection times which are used to calculate the histogram of herald-signal coincidence events, $G_{\rm h,s}^{(2)}(\tau)$, as a function of time delay, $\tau$, between herald and signal detections.
The normalized herald-signal correlation function is calculated as $g_{\rm h,s}^{(2)}=G_{\rm h,s}^{(2)}(\tau)/(r_{\rm h}r_{\rm s}\Delta\tau T)$ where $r_{\rm h,s}$ are the count rates on the herald and signal detectors, $\Delta\tau$ is the histogram bin width and $T$ is the total time over which counts were recorded.
Figure~\ref{fig1}(d) shows the resulting herald-signal correlation function under these conditions with a coupling laser detuned by $\Delta_{\rm c}/2\pi=330$~MHz.

\begin{figure}
	\includegraphics[width=\columnwidth]{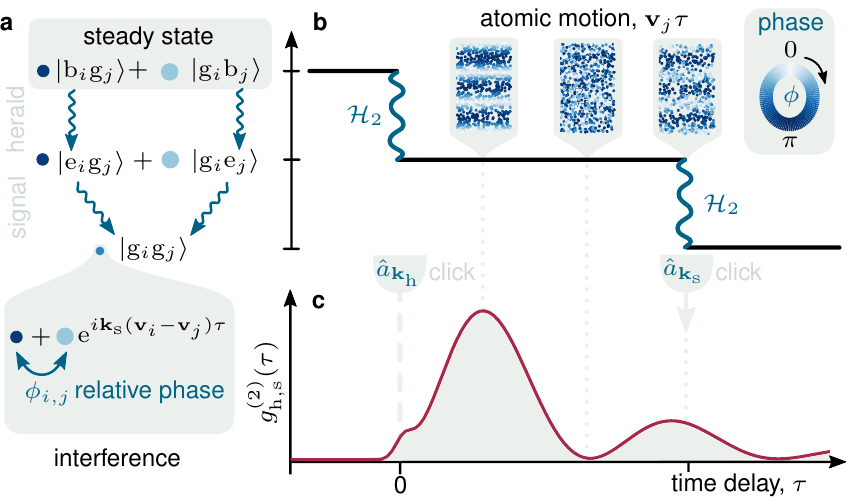}
	\caption{\label{fig2}Collective decay leading to beats.
		(a) Continuous driving prepares the system in a steady state, where atoms $i$ and $j$ (the sum over all possible $i,j$ is implied), are in a superposition of ground $|\rm g\rangle$ and bare $|\rm b\rangle$ states. 
		Herald detection maps the steady-state amplitudes and phases (indicated by the size and color of the circles) into a superposition of excited states $|\rm e\rangle$.
		(b) Subsequently the relative phase of the atoms evolves due to motion (see insets for the spin-wave evolution in space with color coding of the relative phase, the B-field orientation is vertical).
		Since both atoms end up in the same ground state after emitting the signal photon, the emission amplitudes add coherently and a time-dependant factor appears in the collective signal (bottom of (a)).
		(c) This interference leads to beats in the probability of directional signal photon emission over time, $\tau$.}
\end{figure}

\textit{Theoretical model} --- 
The probability of detecting a signal photon a time $\tau$ after heralding, depends on the initial relative phase of the two velocity groups and the speed difference in the signal detector direction.
To understand the process that sets the initial relative phase, and subsequent phase evolution of the atomic medium, consider an ensemble of atoms enumerated by $j$ in the basis $\bigotimes_j |\alpha_j , \mathbf{r}_j, \mathbf{v}_j\rangle \bigotimes |\hat{n}_{\mathbf{k}_{\rm h}}\rangle \bigotimes |\hat{n}_{\mathbf{k}_{\rm s}}\rangle $, where $\alpha \in \{\rm g, a, b, e\}$ denotes the atomic state, and $\hat{n}_{\mathbf{k}_{\rm h}, \mathbf{k}_{\rm s}}$ the occupation of the two decay modes corresponding to the herald and signal wavevectors $\mathbf{k}_{\rm h,s}$.
The atomic dynamics, dominated by evolution under strong laser driving and spontaneous decay to all other free modes, brings the system to the stationary state described by the density matrix $\sum_i c_i |\psi \rangle \langle \psi |$ [Fig.~\ref{fig2}(a)].
Cascaded spontaneous four-wave mixing emission, due to the weak coupling $\mathcal{H}_2$ to the herald and signal modes, can be treated as a perturbative correction to the dynamics.
Detection of a herald photon, $\hat{a}_{\mathbf{k}_{\rm h}}$, therefore projects the system state into the collective spin wave
\begin{equation}\label{eq:projectedsw}
   \hat{a}_{\mathbf{k}_{\rm h}} \mathcal{H}_2 |\psi \rangle \propto \sum_j a_j \mathrm{e}^{-i (k_{\rm h} -k_{\rm c}-k_{\rm p} ) z_j} | \ldots {\rm e}_j \ldots  \rangle,
\end{equation}
where $k_{\rm h}$, $k_{\rm c}$ and $k_{\rm p}$ are the herald, coupling and pump mode wavevectors, and factors $a_j$ depend on the atomic velocity $v_{\rm z}$.
Since the herald detection is broadband, the projection is into a state where a single excitation $|{\rm e}_j\rangle$ is in a superposition of different velocity classes. 

During the subsequent time $\tau$, before emission of the signal photon, the phase of the state given by Eq.~(\ref{eq:projectedsw}) will not change, since state $|{\rm e}_j\rangle$ is decoupled from the strong laser driving.
However, the amplitude of this state will be reduced by $\exp(-\gamma \tau)$ due to spontaneous emission to other spatial modes and homogeneous dephasing mechanisms (e.g. collisions with buffer gasses).
Upon decay of $|{\rm e}_j\rangle$ under $\mathcal{H}_2$ [Fig.~\ref{fig2}(b)], detection of the signal photon $ \hat{a}_{\mathbf{k}_{\rm s}}$ is also broadband, and therefore doesn't differentiate between emission from different velocity classes.
Therefore, no which-path information is measured.
Emission from different velocity classes will, due to atomic motion [Fig.~\ref{fig2}(b) insets], have a frequency shift of $k_{\rm s}v_{\rm z}$.
This can give rise to beats in the signal photon detection [Fig.~\ref{fig2}(c)], provided that \emph{no information is left in the medium} about which atom emitted the photon.
All states where the two atoms, labelled $i$ and $j$, are in the superposition of ground and excited state $c_1(t)|{\rm g}_i{\rm e}_j\rangle+c_2(t)|{\rm e}_i{\rm g}_j\rangle$ fulfil that condition, since after cascaded herald and signal emission (time $\tau$ later) they end up in the \emph{same} state $|{\rm g}_i{\rm g}_j\rangle$ where the amplitude shows interference between the two possible paths $c_1(t+\tau)+c_2(t+\tau)$ [bottom of Fig.~\ref{fig2}(a)].
From this consideration we see that the initial phase of the signal emission from the velocity class $v_z$ will be set by the stationary value of the single-atom coherence element $\rho_{\rm bg}(v_z)$ between the states $|\rm b\rangle$ and $|\rm g\rangle$ for the corresponding velocity.
Integrating over all the velocity classes, weighted according to their probabilities given by the Maxwell-Boltzmann distribution $f(v_{\rm z})$, one obtains (see supplemental material) the two photon correlation function  $\langle \hat{a}_{\mathbf{k}_{\rm s}}^\dagger \hat{a}_{\mathbf{k}_{\rm s}}  \hat{a}_{\mathbf{k}_{\rm h}}^\dagger \hat{a}_{\mathbf{k}_{\rm h}} \rangle_\tau = |\Psi|^2$ where
\begin{equation}\label{evolution}
\Psi \propto \int_{v_{\rm z}} \mathrm{d}v_{\rm z}~f(v_{\rm z})\rho_{\rm bg}(v_{\rm z})\exp [-(\gamma+ik_{\rm s}v_{\rm z})\tau].
\end{equation}

We note that this calculation only includes the contribution from correlated decays and ignores the background of uncorrelated photon counts produced by other events (see supplemental material).
This gives the normalized joint-detection probability for the herald and signal photons, as defined by Glauber's theory~\cite{Glauber1963}, $g^{(2)}_{\rm h,s}(\tau)=1+c|\Psi|^2$, where the constant of proportionality $c$ accounts for the uncorrelated background and is included as a free parameter in the model. 

\begin{figure}
	\includegraphics[width=\columnwidth]{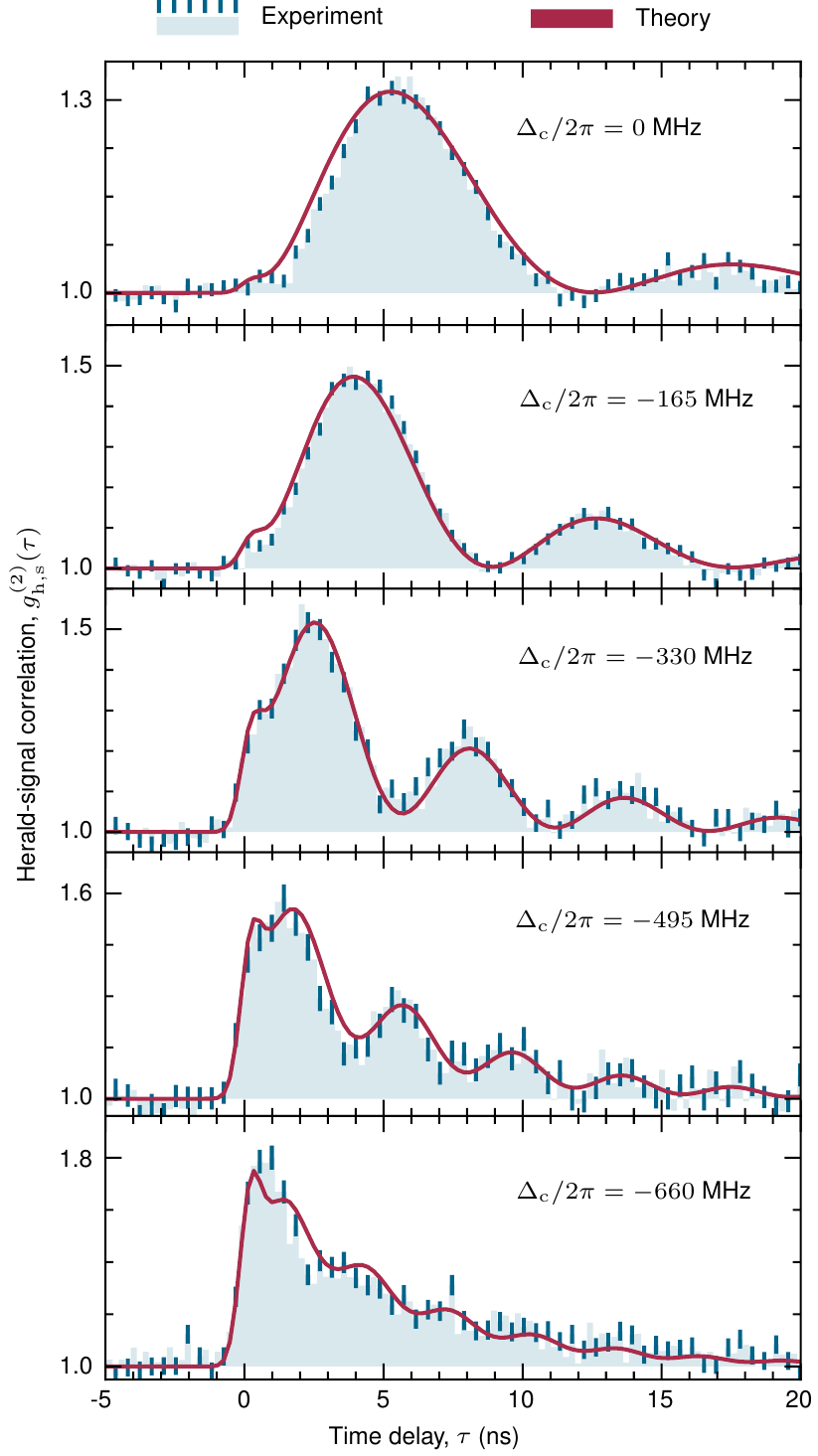}
	\caption{\label{fig3}Persistence of coherence between two collective excitation components. 
		Experimental data (blue) showing interference resulting from coherent splitting of a single excitation across two groups of atoms with relative motion.
		The Doppler shift leads to beats in the state readout with a frequency proportional to the relative velocity.
		The detuning of a strong dressing laser, $\Delta_{\rm c}$, sets the velocities of the excited atoms and thereby determines the beat frequency.
		A theoretical model (red) finds excellent agreement with the data across the entire range of detunings studied.
		The error bars on the experimental data are calculated assuming Poissonian noise on the individual histogram bins~\cite{Hughes2010}.}
\end{figure}

\textit{Experimental results} --- 
The developed theoretical model agrees very well with the temporal correlation data over a wide range of parameters~(Fig.~\ref{fig3}).
The model is fit to the data using chi-squared minimization~\cite{Hughes2010} with common fit parameters for all data-sets displayed.
This agreement demonstrates the excellent understanding and control of the state preparation achieved in our experiment and compares very favourably to the cases without control over the initial state, like recent experiments in pulse-seeded four-wave mixing \cite{Huber2014,Ripka2016}. The observed lifetime of the collective coherence is on the order of the excited state lifetime. During this coherence time, atoms in different velocity groups can be independently perturbed by external fields, e.g. by exploiting their Doppler shifted optical resonances with coherent driving. An applied perturbation would imprint a different phase on the excitation stored in each velocity group, which could be directly measured by the accompanying change in the herald-signal correlation. 
In future an inverted-Y scheme could be used, combining a typical $\Lambda$ scheme with an additional laser that strongly dresses the intermediate state~\cite{Wen2008}. This would enable storage and deterministic retrieval of the split single-photon, due to the long lived ground-state coherences and longer spin-wave period~\cite{Dudin2013}. During the storage time the usual qubit rotation operations could be performed by applying off-resonant driving that imprints a relative phase via the AC-Stark shift. 

The heralded single-photon has primarily two frequency components, the frequencies and amplitudes of which are tunable via the coupling laser parameters and the magnetic field. Such a two-color photon may be a useful resource for entangling two spatially separated atomic quantum memories. In such a scheme each memory would absorb one part of the two-color photon. A symmetric resource state can be prepared by resonant driving ($\Delta_{\rm c}=0$) that symmetrically excites two velocity classes, moving in opposite directions with velocities $\pm \Omega_{\rm c}/(k\sqrt{8})$ set by the coupling laser power through $\Omega_{\rm c}$. Similar diamond schemes in Rubidium would allow for the generation of telecoms-wavelength single photons~\cite{Willis2011}.

In conclusion, excellent agreement between theory and experiment demonstrates that atoms in strongly-dressed thermal vapors~\cite{Sibalic2016} offer a reliable platform for quantum state engineering.
The addition of external magnetic fields allows for selective excitation and observation of well-defined simple systems that can be completely and accurately modelled~\cite{Whiting2015,Whiting2016}.
Collective excitation of two velocity groups is an example of an entangled state that is robust against single atom loss and dephasing~\cite{Dur2000}.
With the emission of two-color heralded single photons providing a direct relative phase measurement, and tunability of the atomic response through adjustments to the dressing laser, these states can be further explored in protocols for quantum state control of atoms and light.

\begin{acknowledgments}
We thank K. J. Weatherill, E. Bimbard, R. S. Mathew and H. Busche for their helpful and informative comments. 
We acknowledge financial support from EPSRC (grant EP/L023024/1) and Durham University.
CSA is supported by the EU project H2020-FETPROACT-2014 184 Grant No. 640378 (RYSQ).
The datasets generated during and/or analysed during the current study are available in the Durham University Collections repository, \url{http://dx.doi.org/10.15128/r19c67wm81t}.
\end{acknowledgments}


%

\clearpage
\onecolumngrid

\section*{\large Supplementary Information}
\setcounter{section}{0}
\setcounter{equation}{0}
\setcounter{figure}{0}

\bigskip
This supplemental information is divided into four sections.
Firstly we present a formal derivation of the herald-signal correlation function [Eq. (2), main text].
Secondly we show the atomic energy level structure in the large magnetic field and discuss the possible spontaneous decay paths.
Thirdly we present additional data regarding the non-classicality of our heralded single photons.
Finally we derive the condition for two-photon absorption resonances in the medium.

\renewcommand\thefigure{S\arabic{figure}}

\section{Derivation of the theoretical model describing the observed beats in the $g_{\rm h,s}^{(2)}$ measurement}

In the following we derive a theoretical prediction for quantum beats in four-wave mixing (FWM) emission due to atomic motion in a single spin-wave excitation.
We calculate the herald-signal joint-detection expectation value $\langle \hat{E}_{\rm s}^\dagger (t+\tau) \hat{E}_{\rm s} (t+\tau) \hat{E}_{\rm h}^\dagger (t) \hat{E}_{\rm h} (t) \rangle$ for a spatially extended atomic ensemble, where $\hat{E}_{\rm h \ldots \rm s}^\dagger (t) \hat{E}_{\rm h \ldots \rm s} (t)$ is the photon number in the herald and signal channels respectively, at time $t$.

Consider the dynamics of an ensemble of $N$ four-level atoms, enumerated with $j$, located at $\mathbf{r}_j$ and moving with velocities $\mathbf{v}_j$, coupled to electromagnetic field (EM) modes (Fig.~\ref{fig1}). 
\begin{figure}[b]
\includegraphics[width=0.5\columnwidth]{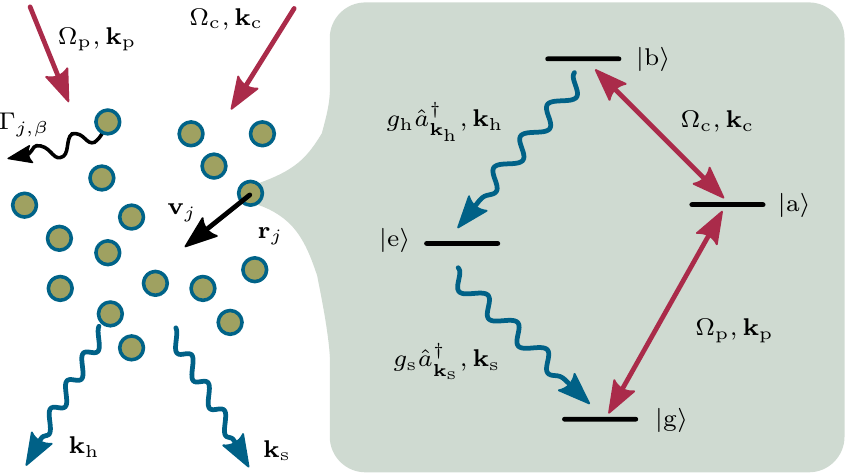}
\caption{\label{fig1}A spatially extended medium ($\mathrm{max}[ |\mathbf{r}_i-\mathbf{r}_j |]\gg2\pi/\mathbf{k}_{\rm s}$) containing $N$ atoms, enumerated by $j$, located at $\mathbf{r}_j$, and moving with velocities $\mathbf{v}_j$. Internally (inset on right) the atoms have four energy levels, and are driven by pump and coupling fields with Rabi frequencies $\Omega_{\rm p}$ and $\Omega_{\rm c}$. Atoms can decay to the herald mode $\mathbf{k}_{\rm h}$ and the signal mode $\mathbf{k}_{\rm s}$ under the influence of $g_{\rm h} \hat{a}_{\mathbf{k}_{\rm h}}$ and $g_{\rm p}\hat{a}_{\mathbf{k}_{\rm s}}$, or to one of the other modes $\beta$ with rate $\Gamma_{j,\beta}$. The system is analysed in the basis $\bigotimes_j |\alpha_j , \mathbf{r}_j, \mathbf{v}_j\rangle \bigotimes |\hat{n}_{\mathbf{k}_{\rm h}}\rangle \bigotimes |\hat{n}_{\mathbf{k}_{\rm s}}\rangle $, $\alpha \in \{\rm g,a,b,e\}$, which is coupled to the Markovian bath of all other vacuum modes.}
\end{figure}
Two of these modes are strong pump and coupling laser fields that will be treated as classical driving fields, whose driving strength is given by Rabi frequencies $\Omega_{\rm p}$ and $\Omega_{\rm c}$, and direction by the wavevectors $\mathbf{k}_{\rm p}$ and $\mathbf{k}_{\rm c}$. 
The dynamics of two field modes named the herald and signal modes, with energies corresponding to the $|\rm b\rangle \rightarrow |\rm e\rangle$ and $|\rm e\rangle \rightarrow |\rm g\rangle$ transitions, are considered separately.
Their spatial directions, labelled by the wavevectors $\mathbf{k}_{\rm h}$ and $\mathbf{k}_{\rm s}$ respectively, are defined by the directions of the single-mode inputs of the single-photon detectors used for the detection of herald and signal photons in the experiment.
All of the empty EM modes, except for the herald and signal modes, will be treated with the usual coupling to the Markovian reservoir, giving rise to spontaneous emission $\Gamma_{j, \alpha}$.
The system is analysed in the basis $\bigotimes_j |\alpha_j , \mathbf{r}_j, \mathbf{v}_j\rangle \bigotimes |\hat{n}_{\mathbf{k}_{\rm h}}\rangle \bigotimes |\hat{n}_{\mathbf{k}_{\rm s}}\rangle $, $\alpha \in \{\rm g,a,b,e\}$.
The dynamics of the internal degrees of freedom are described with the Hamiltonian $\bar{\mathcal{H}} = \bar{\mathcal{H}}_1+\bar{\mathcal{H}}_2$ ($\hbar=1$), where
\begin{eqnarray*}
\bar{\mathcal{H}}_1 &=& \sum_j \left[ \omega_{\rm a} |{\rm a}_j\rangle\langle {\rm a}_j| + \omega_{\rm b} |{\rm b}_j\rangle\langle {\rm b}_j| + \omega_{\rm e} |{\rm e}_j\rangle\langle {\rm e}_j|\right]\\
& & +\sum_j \left[\frac{\Omega_p}{2}~ \mathrm{e}^{i \mathbf{k}_{\rm p} \mathbf{r}_j(t) -i \omega_{\rm p} t }~|{\rm a}_j
\rangle\langle {\rm g}_j|+\frac{\Omega_c}{2}~ \mathrm{e}^{i \mathbf{k}_{\rm c} \mathbf{r}_j(t) -i \omega_{\rm c} t }~ |{\rm b}_j\rangle \langle {\rm a}_j|+h.c.\right]
\end{eqnarray*}
describes the four level system driven, in the rotating wave approximation (RWA), by strong pump and coupling laser fields with respective frequencies $\omega_p$ and $\omega_c$, driving the transitions $|{\rm g}\rangle \leftrightarrow |{\rm a}\rangle$ and $|{\rm a}\rangle \leftrightarrow |{\rm b}\rangle$.
The energies of the states $|\alpha\rangle$ are $\omega_\alpha$.
Additionally,
\begin{eqnarray*}
\bar{\mathcal{H}}_2 &=& \sum_j \left[g_{\rm h}~ \mathrm{e}^{-i \mathbf{k}_{\rm h} \mathbf{r}_j+i\omega_{\rm h} t}~ \hat{a}_{\mathbf{k}_{\rm h}}^\dagger~ |{\rm b}_j\rangle \langle {\rm e}_j| + g_{\rm s}~ \mathrm{e}^{-i \mathbf{k}_{\rm s} \mathrm{r}_j +i \omega_{\rm s}t} ~\hat{a}_{\mathbf{k}_{\rm s}}^\dagger~|{\rm g}_j\rangle\langle {\rm e}_j|+h.c. \right]
\end{eqnarray*}
describes the coupling of the atom, in the RWA, to the herald and signal detection modes.
The coupling strengths between the atom and the vacuum modes, $g_{\rm h}$ and $g_{\rm s}$ for herald and signal channels respectively, formally correspond to $g_{\rm s} = \sum_{\mathbf k \in \mathbf{k}_{\rm s} \pm \Delta \mathbf{k}} g_{\rm be}$ where $|\Delta \mathbf{k}|\ll |\mathbf{k}|$ defines the range of emitted photon directions that hit the detector's sensitive area, and $g_{\rm be}$ is the vacuum Rabi coupling frequency.
The atom coupling to all other modes is described by the Lindblad super-operator $L[\hat{\rho}_N] = \sum_{j,\beta} (L_{j,\beta} \hat{\rho}_N L_{j,\beta}^\dagger - \frac{1}{2}L_{j,\beta}^\dagger L_{j,\beta} \hat{\rho}_N - \frac{1}{2}\hat{\rho}_N L_{j,\beta}^\dagger L_{j,\beta})$, where $L_{j,\beta}$ are the decay channels of atom $j$, enumerated by $\beta$.
Since the coupling of the atom to the herald and signal modes, described by $\mathcal{H}_2$, is negligible compared with the coupling to all the other spatial modes, the decay of states $|\rm b\rangle$ and $|\rm e\rangle$ is still described, to an excellent approximation, by the usual spontaneous decay rates $\Gamma_{\rm b}$ and $\Gamma_{\rm e}$.
Evolution of the external degrees of freedom, due to atomic motion, is accounted for by $\mathbf{r}_j(t) = \mathbf{r}_j(0) + \mathbf{v}_j t$.

Before solving the dynamics, we choose a convenient basis by applying the unitary transformation
\[\hat U = \exp\left( i\sum_j \left\{ [\omega_{\rm p}t - \mathbf{k}_{\rm p}(\mathbf{r}_j(0)+\mathbf{v}_j t)]  |{\rm a}_j \rangle \langle {\rm a}_j|+ [(\omega_{\rm p}+\omega_{\rm c})t - (\mathbf{k}_{\rm p}+\mathbf{k}_{\rm c})( \mathbf{r}_j(0)+\mathbf{v}_j t)]  |{\rm b}_j\rangle \langle {\rm b}_j| +\omega_{\rm e} t |{\rm e}_j\rangle \langle {\rm e}_j| \right\} \right),\]
such that a new evolution Hamiltonian $\mathcal{H}_1+\mathcal{H}_2 = \hat{U} \bar{\mathcal{H}}\hat{U}^\dagger +i \frac{\mathrm{d} \hat{U}}{\mathrm{d} t}\hat{U}^\dagger$ is obtained.
Thus
\begin{eqnarray}
\mathcal{H}_1 &=& \sum_j \left[ -\Delta_1 |{\rm a}_j\rangle\langle {\rm a}_j| - \Delta_2 |{\rm b}_j\rangle\langle {\rm b}_j|\right] +\sum_j \left[\frac{\Omega_p}{2}~|{\rm a}_j
\rangle\langle {\rm g}_j|+\frac{\Omega_c}{2}~  |{\rm b}_j\rangle \langle {\rm a}_j|+h.c.\right], \label{eq:h1}\\
\mathcal{H}_2&=& \sum_j \left\{ g_{\rm h}~ \mathrm{e}^{-i (\mathbf{k}_{\rm h} -\mathbf{k}_{\rm p}- \mathbf{k}_{\rm c})\mathbf{r}_j(0)+i[\omega_{\rm h}+\omega_{\rm e}-\omega_{\rm p} - \omega_{\rm c} + (\mathbf{k}_{\rm p}+\mathbf{k}_{\rm c}-\mathbf{k}_{\rm h})\mathbf{v}_j] t}~ \hat{a}_{\mathbf{k}_{\rm h}}^\dagger~ |{\rm e}_j\rangle \langle {\rm b}_j|\right. \nonumber  \\
& & \left.+ g_{\rm s}~ \mathrm{e}^{-i \mathbf{k}_{\rm s} \mathbf{r}_j +i (\omega_{\rm s}-\omega_{\rm e})t} ~\hat{a}_{\mathbf{k}_{\rm s}}^\dagger~|{\rm g}_j\rangle\langle {\rm e}_j|+h.c. \right\}, \label{eq:h2}
\end{eqnarray}
where $\Delta_1 \equiv \omega_p-\mathbf{k}_{\rm p}\mathbf{v}_j-\omega_{\rm a}$, $\Delta_2 \equiv \omega_p+\omega_c-(\mathbf{k}_{\rm p}+\mathbf{k}_{\rm c})\mathbf{v}_j-\omega_{\rm b}$ are the single and two-photon detunings respectively.

In the following, we are interested in interference effects that originate from two spatially separated locations within the medium and therefore, we solve the dynamics for $N$ atoms in a thermal ensemble.
Since $g_{\rm h}\hat{a}_{\mathbf{k}_{\rm h}}^\dagger, g_{\rm s}\hat{a}_{\mathbf{k}_{\rm s}}^\dagger \ll \Omega_{\rm p},\Omega_{\rm c}$, we treat the dynamics due to $\mathcal{H}_2$ perturbatively.
In the zeroth-order approximation ($\mathcal{H}_2=0$), the system density matrix evolves only under driving $\mathcal{H}_1$ and dissipation $L[\ldots]$.
This is described by the master equation $\frac{\mathrm{d}}{\mathrm{d}t}\hat{\rho}_N = -i [\hat{\rho}_N,\mathcal{H}_1]+L[\hat{\rho}_N] \equiv \mathcal{L}[\hat{\rho}_N]$, which reaches a steady state, $\hat{\rho}_N^{(0)}$, under the Liouvillian $\mathcal{L}$.
The system evolution under $\mathcal{H}_1$ decomposes to the evolution of individual atoms; $\hat{\rho}_N = \bigotimes_j \hat{\rho}_j \bigotimes | 0_{\mathbf{k}_{\rm s}} 0_{\mathbf{k}_{\rm i}} \rangle $, where $\hat{\rho}_j$ is the single-atom density matrix for the $j$-th atom.
In particular, atoms with the same velocity, $\mathbf{v}$, at different spatial locations will evolve under $\mathcal{H}_1$ to the same single-atom density matrix $\hat{\rho}(\mathbf{v})$.
From this it appears that the relative atomic positions are irrelevant. However, we shall see that the relative positions of atoms in the ensemble play a crucial role due to the phase factor in $\mathcal{H}_2$.

In order to obtain the herald-signal joint-detection correlation function $g^{(2)}_{\rm h,s}(\tau)$ we are interested in calculating $\langle \hat{E}_{\rm s}^\dagger (t+\tau) \hat{E}_{\rm s} (t+\tau) \hat{E}_{\rm h}^\dagger (t) \hat{E}_{\rm h} (t)\rangle$.
The first non-zero contribution to this element originates from the second order perturbation by $\mathcal{H}_2$ (Fig.~2(b), main text).
Initially,  $\mathcal{H}_2$ acts on $\hat{\rho}_N^{(0)}$, causing emission of a herald photon at some time $t$.
The system will subsequently evolve under $\mathcal{L}$ and at some time $\tau$ later a signal photon is emitted under the influence of $\mathcal{H}_2(t+\tau)$:
\begin{eqnarray}\label{eq:g2eqn}
\langle \hat{E}_{\rm s}^\dagger (t+\tau) \hat{E}_{\rm s} (t+\tau) \hat{E}_{\rm h}^\dagger (t) \hat{E}_{\rm h} (t) \rangle & = & \Tr\left[  \hat{E}_{\rm s}^\dagger (t+\tau) \hat{E}_{\rm s} (t+\tau) \hat{E}_{\rm h}^\dagger (t) \hat{E}_{\rm h} (t) ~ \hat{\rho}_N^{(2)}\right] , \\
\hat{\rho}_N^{(2)} &=& \mathcal{H}_2 (t+\tau)~ \mathrm{e}^{-i\mathcal{L} \tau}[\mathcal{H}_2(t)~\rho_N^{(0)}~ \mathcal{H}_2^\dagger(t)] ~\mathcal{H}_2^\dagger (t+\tau), \nonumber
\end{eqnarray}
where the trace is over all the atomic degrees of freedom and the herald and signal field modes.

Analysing the time dependence of the atom coupling to the herald mode, i.e. the terms containing $\hat{a}_{\mathbf{k}_{\rm h}}$ in $\mathcal{H}_2$ [Eq.(\ref{eq:h2})], we see that for atoms with a velocity $\mathbf{v}$ the dominant decay is to a mode with frequency $\omega_{\rm h} = \omega_{\rm p}+\omega_{\rm c}-\omega_{\rm e}-(\mathbf{k}_{\rm p}+\mathbf{k}_{\rm c}-\mathbf{k}_{\rm h})\mathbf{v}_j$.
Starting from the steady state density matrix $\hat{\rho}_N^{(0)}$, the emission of a photon in the herald mode acts on the states as
\begin{equation*}
\hat{\rho}_N^{(1)}(t)\equiv \mathcal{H}_2 \hat{\rho}_N^{(0)}  \mathcal{H}^\dagger_2 \propto \sum_i c_i \left[\sum_{j_1} c_{j_1}'~g_{\rm h}~ \textrm{e}^{-i (\mathbf{k}_{\rm h}-\mathbf{k}_{\rm p}-\mathbf{k}_{\rm c}) \mathbf{r}_{j_1}(t)}  |\ldots {\rm e}_{j_1} \ldots 1_\mathbf{k_{\rm h}} \rangle \right] \left[ \sum_{j_2} c_{j_2}'~ g_{\rm h}~ \textrm{e}^{i (\mathbf{k}_{\rm h}-\mathbf{k}_{\rm p}-\mathbf{k}_{\rm c}) \mathbf{r}_{j_2}(t)} \langle \ldots {\rm e}_{j_2} \ldots 1_{\mathbf{k}_{\rm h}} | \right].
\end{equation*}
We see that the emission, and subsequent detection of the signal photon, projects the system into a state where \emph{a single excitation is stored collectively as a coherent spin-wave} with a periodic phase variation given by the wavevector $\mathbf{k}_{\rm h}-\mathbf{k}_{\rm p}-\mathbf{k}_{\rm c}$.
The broadband detection scheme does not discern the frequency of the herald photon $\omega_{\rm h}$, since $\hat{E}_{\rm h} = \sum_{\omega_{\rm h}} \hat{a}_{\mathbf{k}_{\rm h}}$ where the sum over $\omega_{\rm h}$ encompasses the full Doppler-broadened emission profile from the vapour.
Therefore the system will be projected into a state where \emph{the excitation is stored in all atomic velocity classes}.
In the experiment, two narrow velocity-groups provide the dominant contribution to the amplitude of the excitation: one nearly stationary, and the other centred on a non-zero velocity (Fig.~1(b), main text).
Subsequently, during a time $\tau$ the atoms move to new locations $\mathbf{r}_j(\tau) = \mathbf{r}_j(0)+\mathbf{v}_j\tau$.
During this time the internal state of the system changes only due to the atoms in state $|\rm e\rangle$, since all other atoms are already in a stationary state of $\mathcal{L}$.
This state is decoupled from $\mathcal{H}_1$ [Eq.~(\ref{eq:h1})], but evolves due to spontaneous decay and dephasing collisions under $L[\ldots]$, resulting in an amplitude reduction of $\exp(-\gamma \tau)$.
Upon signal photon emission, the system will be left in the state
\begin{eqnarray*}
\hat{\rho}_N^{(2)} \equiv \mathcal{H}_2(\tau)~\hat{\rho}_N^{(1)}(t+\tau) ~\mathcal{H}_2^\dagger(\tau) &\propto & \exp (-2 \gamma \tau)  \\
& & \times \left\{ \sum_{j_1}  \exp[-i(\mathbf{k}_{\rm h}+\mathbf{k}_{\rm s}-\mathbf{k}_{\rm p}-\mathbf{k}_{\rm c})\mathbf{r}_{j_1}(t)+i(\omega_{\rm s}-\mathbf{k}_{\rm s}\mathbf{v}_{j_1}-\omega_{\rm e}) \tau] ~ |\ldots {\rm g}_{j_1} \ldots 1_{\mathbf{k}_{\rm h}} 1_{\mathbf{k}_{\rm s}} \rangle  \right\}\\
& & \times \left\{ \sum_{j_2}  \exp[i(\mathbf{k}_{\rm h}+\mathbf{k}_{\rm s}-\mathbf{k}_{\rm p}-\mathbf{k}_{\rm c})\mathbf{r}_{j_2}(t)-i(\omega_{\rm s}-\mathbf{k}_{\rm s}\mathbf{v}_{j_2}-\omega_{\rm e}) \tau] ~\langle \ldots {\rm g}_{j_2} \ldots 1_{\mathbf{k}_{\rm h}} 1_{\mathbf{k}_{\rm s}} |  \right\} \\
& &+\ldots,
\end{eqnarray*}
where we have explicitly omitted terms that do not contribute to the correlated emission of photons in the herald and signal channels.
In order for this event to have a significant probability of occurring at any time $\tau$, the emitted signal photon must contain frequencies centred on $\omega_{\rm s} = \omega_{\rm e}+\mathbf{k}_{\rm s}\mathbf{v}_{j}$.
In other words, velocity classes differing by $\delta \mathbf{v}$ will emit photons with frequencies differing by $\mathbf{k}_{\rm s}\delta\mathbf{v}$, with \emph{well defined initial relative phases and amplitudes} set by the emission of an initial herald photon.
Crucially, since the signal detector does not discern the close energies of the emitted photons, in calculating the amplitude for the detection event $\hat{E}_{\rm s} = \sum_{\omega_{\rm s}} \hat{a}_{\mathbf{k}_{\rm s}}$ we must sum over the range of $\omega_{\rm s}$ corresponding to the detector bandwidth. In this way we \emph{do not measure which velocity class emitted the photon}.

If the amplitudes of photon emission from different velocity classes are to interfere in time, causing beats in the signal photon detection, photons \emph{must not leave any information in the atomic medium} about which atom stored the excitation.
States that fulfil this condition have atoms $j_1$ and $j_2$ in a coherent superposition where one is excited to $|\rm b\rangle$ and the other is in the ground state $|\rm g\rangle$, i.e. $| \ldots {\rm g}_{j_1}\ldots {\rm b}_{j_2}\ldots \rangle$ and $| \ldots {\rm b}_{j_1}\ldots {\rm g}_{j_2}\ldots \rangle$.
Since after the two-photon decay both of these states end up with both atoms in the ground state $| \ldots {\rm g}_{j_1}\ldots {\rm g}_{j_2}\ldots \rangle$, there is no information left in the medium conveying which of the two atoms decayed.
This leads to interference in the ground state amplitudes, obtained as a sum of decays from different atoms (Fig.~2(a), main text).
Therefore, in calculating Eq.~(\ref{eq:g2eqn}) the dominant non-zero elements~\cite{Note1} will originate from $ \langle \ldots {\rm g}_{j_1} \ldots {\rm b}_{j_2}\ldots |\hat{\rho}_N^{(0)} | \ldots {\rm b}_{j_1} \ldots {\rm g}_{j_2}\ldots\rangle$ and the corresponding conjugate.
Given that the dynamics under $\mathcal{L}$ decompose into the single-atom dynamics, the contributing matrix elements, traced over all atoms other than $j_1, j_2$, are equal to $\hat{\rho}_{\rm gb}(\mathbf{v}_{j_1}) ~\hat{\rho}_{\rm bg}(\mathbf{v}_{j_2})$, where $\hat{\rho}(\mathbf{v})$ is the steady-state single-atom density matrix.
Therefore the initial phase and amplitude of the emission from state $|\rm e\rangle$ is inherited, by the signal emission process, from $\hat{\rho}_{\rm bg}$.

Overall, the joint detection probability [Eq.~(\ref{eq:g2eqn})] can be written as
\begin{eqnarray*}
\langle \hat{E}_{\rm i}^\dagger (t+\tau) \hat{E}_{\rm i} (t+\tau) \hat{E}_{\rm s}^\dagger (t) \hat{E}_{\rm s} (t) \rangle  = \left| \sum_j ~ g_{\rm s} g_{\rm i}~\hat{\rho}_{\rm bg}(\mathbf{v}_j)
~\exp (-\gamma \tau) ~\exp(-i\mathbf{k}_{\rm i}\mathbf{v}_j \tau) ~\exp[i(\mathbf{k}_{\rm p} +\mathbf{k}_{\rm c}-\mathbf{k}_{\rm s}-\mathbf{k}_{\rm i})\mathbf{r}_j(t)]\right|^2 .
\end{eqnarray*}
In order to obtain non-zero values, summation over random atomic positions $\mathbf{r}_j$ must produce a constant value, which gives rise to the condition $\mathbf{k}_{\rm p} +\mathbf{k}_{\rm c}-\mathbf{k}_{\rm s}-\mathbf{k}_{\rm i}=0$, which is the usual \emph{wave matching condition} for wave-mixing processes in extended media.
When this condition is fulfilled, the remaining time dependence can be written as an integral over all velocity classes [c.f. Eq.~(2), main text]
\begin{equation*}
\langle \hat{E}_{\rm i}^\dagger (t+\tau) \hat{E}_{\rm i} (t+\tau) \hat{E}_{\rm s}^\dagger (t) \hat{E}_{\rm s} (t) \rangle \propto \left| \underbrace{\int_\mathbf{v}\mathrm{d}\mathbf{v}~f(\mathbf{v})~ \rho_{\rm bg}(\mathbf{v})\exp(-\gamma \tau)~\exp(-i \mathbf{k}_{\rm i}\mathbf{v} \tau) }_{\equiv \Psi} \right|^2,
\end{equation*}
where $f(\mathbf{v})$ is the probability density function that an atom has velocity $\mathbf{v}$.
We note that this calculation only includes the contribution from correlated decays.
There is also a constant background of uncorrelated decays produced by other events.
For example, following herald emission in channels other than $\mathbf{k}_{\rm h}$ there is no clear phase matching condition for the signal emission, which can then still end up in $\mathbf{k}_{\rm s}$.
Furthermore, in collisional processes population is transferred non-radiatively from $|\rm a\rangle$ to $|e\rangle$, causing additional background emission.
Due to this the normalised signal for detection, with low heralding efficiency, will have the form
\[
\frac{ \langle \hat{E}_{\rm s}^\dagger (t+\tau) \hat{E}_{\rm s} (t+\tau) \hat{E}_{\rm h}^\dagger (t) \hat{E}_{\rm h} (t) \rangle}{ \langle \hat{E}_{\rm s}^\dagger
\hat{E}_{\rm s}\rangle \langle \hat{E}_{\rm h}^\dagger \hat{E}_{\rm h}  \rangle }= 1+c|\Psi|^2,
\]
where $c$ is a constant dependant on the background level.

\vspace{\columnsep}
\vspace{\columnsep}
\twocolumngrid

\section{Application of a large magnetic field}
\begin{figure}
\includegraphics[width=\columnwidth]{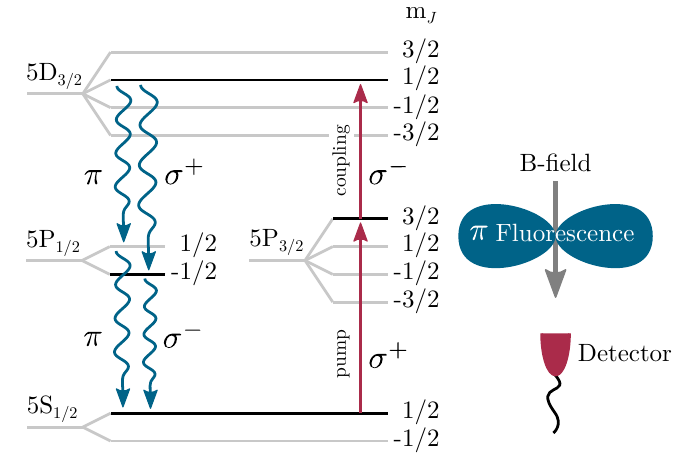}
\caption{\label{quantumbeats}
	A diagram of the atomic energy levels in a large magnetic field and the optical transitions relevant to the experiment.
	The detector only receives light emitted by $\sigma^{\pm}$ atomic transitions due to the applied magnetic field.}
\end{figure}

The internal atomic states are split by the application of a large magnetic field of magnitude 0.6~T.
In this field the electron spin-orbit and nuclear spin angular momenta almost completely decouple, this is called the hyperfine Paschen-Back regime.
The energy eigenstates are the basis states of the $m_{J}$ and $m_{I}$ basis.
The level structure of $^{87}$Rb in the magnetic field is shown in Fig.~\ref{quantumbeats} (without the $m_{I}$ structure for clarity).
In the experiment the magnetic field is aligned with the axis defined by the pump and coupling lasers.
The only transitions that can be driven by the pump and coupling fields in this geometry are $\sigma^{+}$ and $\sigma^{-}$ transitions.
The pump laser is tuned to the $\sigma^{+}$ transition 5S$_{1/2}(m_{J}=1/2)\rightarrow$5P$_{3/2}(m_{J}=3/2)$ and the coupling laser is tuned to the $\sigma^{-}$ transition 5P$_{3/2}(m_{J}=3/2)\rightarrow$5D$_{3/2}(m_{J}=1/2)$.
These transitions are separated from their neighbours in frequency by more than the Doppler-broadened linewidths.
Therefore only the $\left|5D_{3/2},~m_{J} = 1/2\right>$ state is populated by the driving fields.
From here the atoms can decay to the initial ground state via two paths.
Because of the magnetic field the atoms that decay via the $\pi$ transitions cannot emit light in the direction of the detectors.
Furthermore, we apply polarization filtering to the signal mode such that only the light from the $\sigma^{-}$ transition can be detected.
This means photons reaching the detectors must come from a single decay pathway in the atoms and therefore single-atom quantum beats cannot be observed.
Finally, we note that if one chooses a different geometry where light from the $\pi$ transitions can be detected, the single-atom quantum beat frequency will be of the order of 10s of GHz due to the large splitting of the intermediate states.

\section{Non-classical correlations}
\begin{figure}[ht]
\includegraphics[width=\columnwidth]{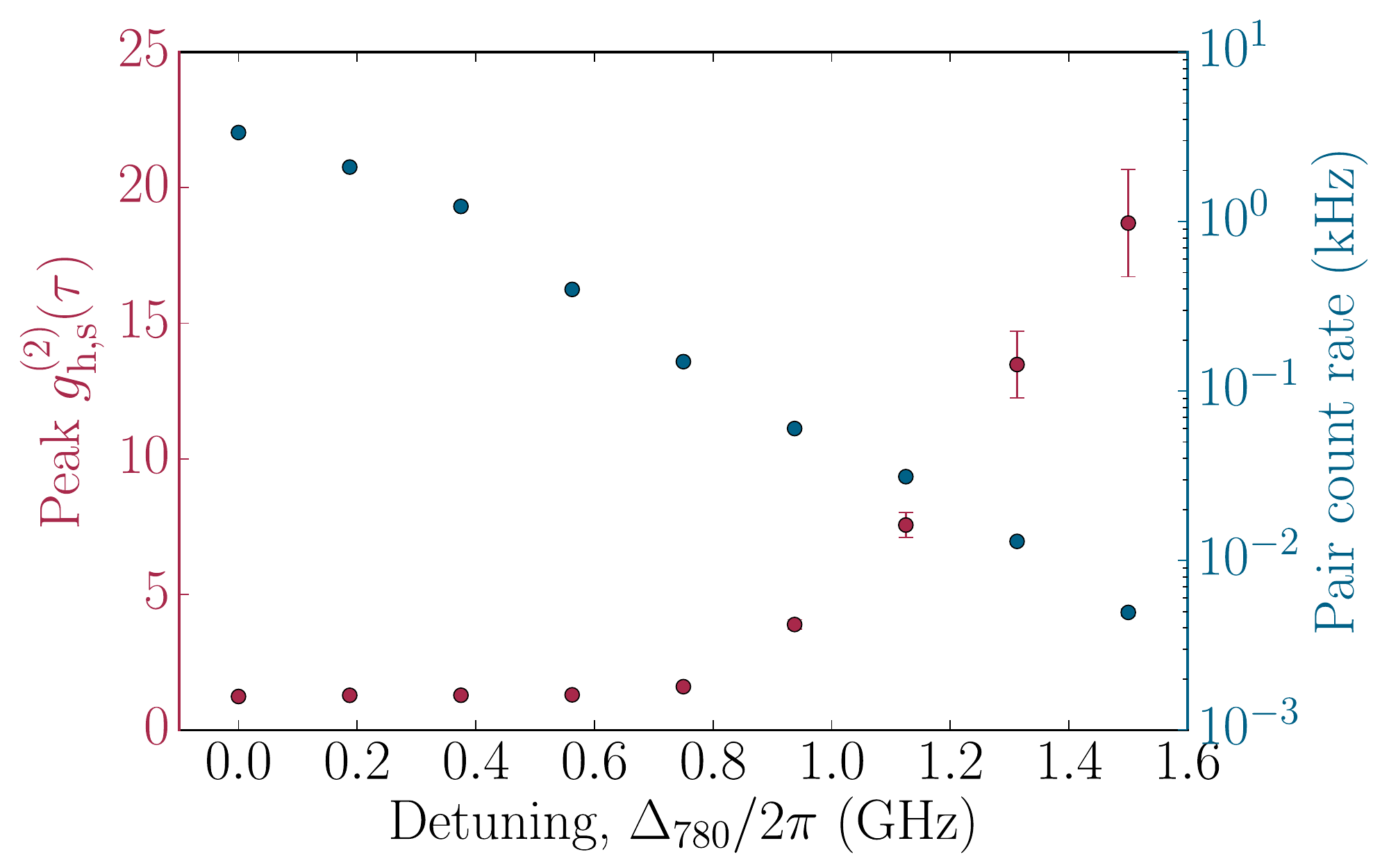}
\caption{\label{det}
	The rate of heralded single photons and the maximum value of the herald-signal correlation function as a function of pump detuning.
	At large detunings the correlation $g_{\rm h,s}^{(2)}(\tau) \gg 2$ which violates the Cauchy-Schwarz inequality, showing the non-classical nature of the photon correlations in our system.}
\end{figure}

In the main text we present herald-signal correlation data for a resonant 780~nm pump laser and a near resonance 776~nm coupling laser (Fig.~3, main text).
These detunings were chosen because the multi-atom quantum beats are most strongly evident in this data.
However, in this regime the correlations do not show a maximum value of $g_{\rm h,s}^{(2)}(\tau)$ that violates the Cauchy-Schwarz inequality; near resonance the background of uncorrelated photons is too large.
For a resonant coupling laser and a detuned pump laser we observe a much larger correlation at the expense of a lower heralded photon rate (Fig.~\ref{det}).

\section{Derivation of the 2-photon absorption resonance condition}
Starting from the interaction Hamiltonian for a three level ladder system interacting with two co-propagating driving fields in the rotating wave approximation,
\begin{equation*}
\hat H_{\rm int}=\frac{\hbar}{2}
\begin{pmatrix}
0 & \Omega_{\rm p} & 0 \\
\Omega_{\rm p} & -2\Delta_{\rm p} & \Omega_{\rm c} \\
0 & \Omega_{\rm c} & -2(\Delta_{\rm p}+\Delta_{\rm c}) \\
\end{pmatrix},
\end{equation*}
we find the three dressed state energies (eigenvalues) in the case of $\Omega_{\rm p}\ll\Omega_{\rm c}$,
\begin{equation*}
E=0,~ \frac{E}{\hbar}=-\left(\Delta_{\rm p}+\frac{\Delta_{\rm c}}{2}\right) \pm \frac{1}{2}\sqrt{\Delta_{\rm c}^{2}+\Omega_{\rm c}^{2}}.
\end{equation*}
A 2-photon absorption resonance occurs when the dressed states are on resonance with the driving fields, i.e. when the dressed state energies are zero.
This is true for
\begin{equation*}
\Delta_{\rm p}^{2}+\Delta_{\rm p}\Delta_{\rm c} = \frac{\Omega_{\rm c}^{2}}{4}.
\end{equation*}
Including the Doppler shift $\Delta_{\rm p,c}\rightarrow\Delta_{\rm p,c}-k_{\rm p,c}v_{\rm z}$ and setting $\Delta_{\rm p}=0$ as in the experiment, we can write
\begin{equation*}
(k_{\rm p}^{2}+k_{\rm p}k_{\rm c})v_{\rm z}^{2} - \Delta_{\rm c}k_{\rm p}v_{\rm z} - \frac{\Omega_{\rm c}^{2}}{4} = 0
\end{equation*}
which we can solve to find the velocity classes for which the 2-photon absorption resonance condition is met,
\begin{equation*}
v_{\rm z} = \frac{\Delta_{\rm c}k_{\rm p}}{a} \pm \sqrt{\frac{\Delta_{\rm c}^{2}k_{\rm p}^{2}}{a^{2}}+\frac{\Omega_{\rm c}^{2}}{2a}}
\end{equation*}
where $a = 2(k_{\rm p}^{2}+k_{\rm p}k_{\rm c})$.
In our experiment the excitation states are nearly equi-spaced hence setting $k_{\rm p} = k_{\rm c} = k$ we arrive at
\begin{equation*}
v_{\rm z} = \frac{1}{4k}(\Delta_{\rm c}\pm \sqrt{\Delta_{\rm c}^{2}+2\Omega_{\rm c}^{2}}).
\end{equation*}

\end{document}